\documentstyle[sprocl,epsfig]{article}
\newcommand{\be}{\begin{eqnarray}}
\newcommand{\ee}{\end{eqnarray}}
\def\e{\epsilon}

\def\be{\begin{equation}}
\def\ee{\end{equation}}
\def\bea{\begin{eqnarray}}
\def\eea{\end{eqnarray}}

\begin{document}

\begin{flushright}
{\bf KAIST--TH--98/04}
\end{flushright}

\vskip 1.0cm

\title{
New Electroweak Tests for the Topflavour Model
\footnote{{\it Talk delivered at the first APCTP Workshop
on Pacific Particle Physic Phenomenology, 
Oct. 31 -- Nov. 2, 1997, Seoul, Korea}}
}

\author{ Jong Chul Lee and
Kang Young Lee\footnote{conference speaker} }
\address{
Department of Physics, 
Korea Advanced Institute of Science and Technology\\
Taejon 305 -- 701, Korea
}

\maketitle
\abstracts{
We explore phenomenologies of the topflavour model
for the LEP experiment at $m_{_Z}$ scale.
Implications of the model on the $Z$ peak data are studied
in terms of the precision variables $\epsilon_i$'s.
}

\section{Introduction}

In this work we consider the model with additional SU(2)
acting only on the third generation,
which has been suggested by several authors \cite{malkawi,nandi}
as a possible solution of the $R_b$ problem.
The third generation undergoes different flavour dynamics
from the first and second generations
and we expect that this type of model would help us to explain
the discrepancy of $R_b$ of which measurement is about 1.8 (LEP)
and 1.5 (LEP+SLC) standard deviations higher than the SM prediction.
As an analogy to the topcolor model \cite{topcolor},
this model is called topflavour model \cite{nandi}.

In order to parametrize new physics effects on the observables
from the LEP experiments, we calculate the precision
variables $\epsilon_i$'s introduced by Altarelli et al.
\cite{altarelli2,altarelli1} 
with the new LEP data reported by the LEP Electroweak Working Group
\cite{LEPEWWG} in the framework of the topflavour model.
Because there is no direct evidence for new physics beyond the 
Standard Model (SM) at LEP until now, 
the new physics contribution, if it exists,
are thought to be comparable with the radiative correction
effects of the SM at most.
Hence it will be interesting to study the new physics effects
in terms of the precision variables.
Among four epsilon variables, $\epsilon_b$ has been of particular interest
because it encodes the corrections to the $Z \to b \bar b$ vertex
and we focus on $\epsilon_b$.

\section{The Epsilon Variables and the Standard Model Predictions }

The precision variables 
$\e_1$, $\e_2$, $\e_3$ and $\e_b$ are defined from the basic observables,
the mass ratio of $W$ and $Z$ bosons $m_W/m_Z$, the leptonic
width $\Gamma_l$, the forward--backward asymmetry for
charged leptons $A_{FB}^l$ and the $b$--quark width $\Gamma_b$,
which are all defined at the $Z$--peak.
In terms of the epsilon variables,
we have the virtue that the most interesting physical results are 
already obtained at a completely model independent manner
without assumptions like the dominance of vacuum polarisation diagrams.
The $m_t$--dependence for all observables
via loops come out through the $\epsilon_i$'s.

The $\epsilon_{1,2,3}$ variables are defined by 
the linear combinations of the correction terms 
$\Delta \rho$, $\Delta k$ and $\Delta r_{_W}$ 
in the eqs. (9) of ref. [5], 
which are extracted from the vector and axial--vector couplings 
for charged leptons, $g_{_{lV}}$, $g_{_{lA}}$ 
and the mass ratio of $W$ and $Z$ bosons, $m_W/m_Z$.
The vector and axial--vector couplings are obtained 
from the observables $\Gamma_l$ and $A_{FB}^l$.
The formular for $\epsilon_b$ is rather complicated. 
It is defined by the equations :
\be
g_{_{bA}} = -\frac{1}{2} \left( 1+\frac{\Delta \rho}{2} \right)
          (1+\epsilon_b),~~~~
\frac{g_{_{bV}}}{g_{_{bA}}} 
     = \frac{ 1-\frac{4}{3}(1+\Delta k) s_0^2 + \epsilon_b}
            {1+\epsilon_b}~~.
\ee
We obtain the relation between $\epsilon_b$ and $\Gamma_b$ by 
insertion of $g_{_{bV}}$ and $g_{_{bA}}$ into the formular
of $\Gamma_b$.
The SM predictions for $\epsilon_i$'s are given by
\be
\epsilon_1 = 5.8 \times 10^{-3},~~
\epsilon_2 = -7.4 \times 10^{-3},~~
\epsilon_3 = 5.0 \times 10^{-3},~~
\epsilon_b = -5.3 \times 10^{-3},
\ee
with the Higgs boson mass $m_{_H} = 100$ GeV.

The epsilon variables are obtained using the recent LEP data 
from ref. [6]:
\begin{eqnarray}
\epsilon_1 = (2.9 ~~\sim~~ 5.3) \times 10^{-3}&,&
\epsilon_2 = (-6.6 ~~\sim~~-3.6) \times 10^{-3}~~,
\nonumber \\
\epsilon_3 = (1.0 ~~\sim~~ 4.5) \times 10^{-3}&,&
\epsilon_b = (-4.5 ~~\sim~~-0.8) \times 10^{-3}~~.
\end{eqnarray}
Note that the lepton universality is assumed 
for the values of $\Gamma_l$ and $A_{FB}^l$.

\section{ The Topflavour Model}

We study the topflavour model 
with the extended electroweak gauge group  
$SU(2)_l \times SU(2)_h \times U(1)_Y $  where 
the first and second generations couple to $SU(2)_l$ 
and the third generation couples to $SU(2)_h$.
The left--handed quarks and leptons in the first and second generations
transform as (2,1,1/3), (2,1,-1) under 
$SU(2)_l \times SU(2)_h \times U(1)_Y $,
and those in the third generation as (1,2,1/3), (1,2,-1)
while right--handed quarks and leptons transform as (1,1,2$Q$)
where $ Q = T_{3l} + T_{3h} + Y $ 
is the electric charge of a fermion.
 
The covariant derivative is given by 
$ D^{\mu} = \partial^{\mu} + i g_l T^{a}_{l} W^{\mu}_{la} 
          + i g_h T^{a}_{h} W^{\mu}_{ha} 
          + i g^{\prime} Y B^{\mu} $, 
where $ T^{a}_{l}$ and $T^{a}_{h}$ denote the $SU(2)_{(l,h)}$ 
generators and $Y$ is the $U(1)$ hypercharge generator. 
Corresponding gauge bosons are $ W^{\mu}_{la} , W^{\mu}_{ha} $
and $B^{\mu}$ with the coupling constants  
$g_{l}, g_{h}$ and $g^{\prime}$ respectively.
The gauge couplings may be written as
$
g_l = \frac{e}{\sin \theta \cos \phi} ,\mbox{~~~~}
g_h = \frac{e}{\sin \theta \sin \phi},\mbox{~~~~} 
g^{\prime} = \frac{e}{\cos \theta} 
$
in terms of the weak mixing angle $\theta $ and the new mixing 
angle $\phi$ between $SU(2)_l$ and $SU(2)_h$.

The symmetry breaking is accomplished  by the vacuum expectation values 
(VEV) of two scalar fields  $\Sigma$ and $\Phi$:
$ \langle \Phi \rangle = ( 0, v/\sqrt{2} )^{\dagger}$,
$ \langle \Sigma \rangle = u I$ where $I$ is $2\times2$
identity matrix.
The scalar field $\Sigma $ transforms as (2,2,0) 
under $ SU(2)_l \times SU(2)_h \times U(1) $
and we choose that $\Phi$ transforms as (2,1,1) corresponding to 
the SM Higgs field.   
In the first stage, 
the scalar field $\Sigma$ gets the vacuum expectation value and
breaks $SU(2)_l \times SU(2)_h \times U(1)_Y $ down 
to $ SU(2)_{l+h} \times U(1)_Y $ at the scale $\sim u$.
The remaining symmetry is broken down to $U(1)_{em}$  
by the the VEV of $\Phi$ at the electroweak scale. 
Since the third generation fermions do not couple to Higgs fields
with this particle contents,
they should get masses via higher dimensional operators.
We do not study the mass generation problem of this model
in details here.
We demand that both $SU(2)$ interactions are perturbative 
so that the value of the mixing angle $\sin \phi$ 
is constrained $g_{(l,h)}^2/4 \pi < 1$,
which results in $ 0.03 < \sin^2 \phi < 0.96 $.
We also assume that the first symmetry breaking scale is much higher than
the electroweak scale, $ v^2/u^2 \equiv \lambda \ll 1 $.

The basic observables in the topflavour model depend upon
the model parameters $\lambda$ and $\sin^2 \phi$
as well as the Higgs mass $m_{_H}$.
With the correction terms we obtain the epsilon variables
in the topflavour model \cite{leelee}:
\begin{eqnarray}
\epsilon_1 &=& \epsilon^{SM}_1 - 2 \lambda \sin^4 \phi
\nonumber \\
\epsilon_2 &=& \epsilon^{SM}_2 +  \lambda \sin^4 \phi 
                             \left( \frac{s_0^2}{c_0^2 - s_0^2} -2
                             \right)
\nonumber \\
\epsilon_3 &=& \epsilon^{SM}_3 - \lambda \sin^4 \phi
\nonumber \\
\epsilon_b &=& \epsilon^{SM}_b - \frac{1}{2} \lambda 
                        \frac{\sin^2 \phi}{g_{lA}^{SM}}
                        (1+\epsilon_b^{SM} \sin^2 \phi)~~.
\end{eqnarray}
We used the ZFITTER \cite{ZFITTER} for numerical
calculations of the epsilon variables in this model
and we use 175 GeV as input value of $m_t$, 
which is reported by the CDF and D0 \cite{top}.

\begin{figure}[h]
\label{figone}
\epsfig{file=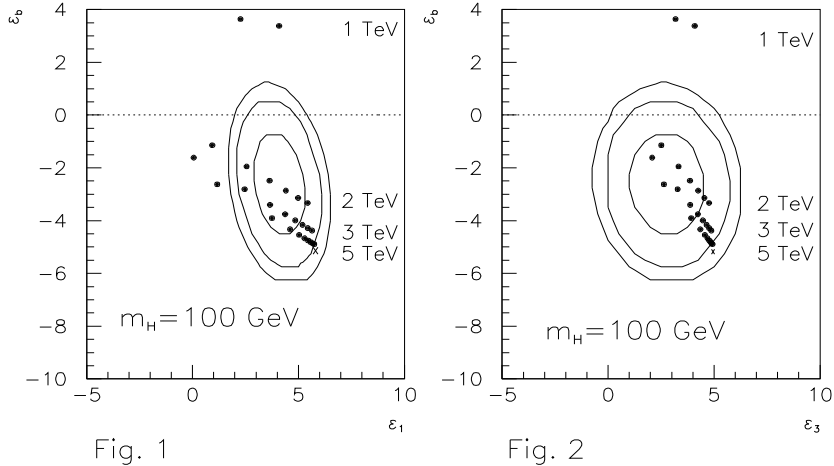,height=8.0cm,width=13.0cm}
\begin{flushleft}
\caption{Plots of the model predictions in units of $10^{-3}$
with varying the model parameter
$sin^2 \phi$, $m_{_{Z'}}$ and the Higgs boson mass $m_H$ in
$\epsilon_1$--$\epsilon_b$ plane.
The experimental ellipses at 1-$\sigma$, 90 \% C.L.
and 95 \% C.L. are given.
Points are labelled from $\sin^2 \phi = 0.1~ \sim ~0.9$
from right to left for each value of $m_{_{Z'}}$.}
\caption{Plots of the model predictions in units of $10^{-3}$
with varying the model parameter
$sin^2 \phi$, $m_{_{Z'}}$ and the Higgs boson mass $m_H$ in
$\epsilon_3$--$\epsilon_b$ plane.
}
\end{flushleft}
\end{figure}

In Figs. 1 and 2, the experimental ellipses 
are shown in the $\epsilon_1-\epsilon_b$ 
and $\epsilon_3-\epsilon_b$ planes respectively
with the results of the SM and the topflavour model. 
We express the results with variations of
$\sin^2 \phi$ and $m_{_{Z'}}$
where
\be
m_{Z'}^2 = \frac{m_Z^2}{\lambda \sin^2 \phi \cos^2 \phi}
           \left[ \cos^2 \theta_{SM} + \lambda \sin^4 \phi
                  \left( \frac{c_0^2 s_0^2}{c_0^2-s_0^2} + 2 c_0^2
                  \right)
           \right]~~,
\ee
with 
the Weinberg angle of the SM, $\theta_{SM}$. 
The ellipses are shifted along the minus direction of
$\epsilon_b$ compared with those from 1996 data \cite{kylee,parklee}
and thus come closer to the SM prediction.
We find that the SM predictions still lie
outside the 90 \% C.L. ellipses in both planes.
These deviations are caused by 
the $Z b \bar{b}$ coupling and related to the $R_b$ discrepancy.
The lower limit of the mass of heavy gauge boson $Z'$
is about 1.2 TeV (1.1 TeV) at 90 (95) \% C. L.
which agrees with the result of ref. [1].

\section{Concluding Remarks}

In this work, we explore the phenomenologies of the topflavour model, 
which are extension of the SM with the additional $SU(2)$ symmetry.
The third generation is special in this model
and it is also expected to explain the heierarchy of the fermion
mass spectrum as well as the $R_b$ discrepancy observed
in LEP experiment.
The experimantal values of $\epsilon$ variables are obtained
from the recent LEP results reported 
by the LEP Electrweak Working Group.
At present the SM predictions is still out of 90 \% C. L. ellipses 
mainly caused by the deviations of $\epsilon_b$
originated by the $Z b \bar{b}$ vertex.
Since the topflavour model provides the different flavour dynamics
on the third generations, the value of $\epsilon_b$ can shift
to the experimental value.

In conclusion, we investigated the phenomenological implications
of the topflavour model with the LEP data at $Z$--peak.
We found that the best fitted mass of $Z'$ boson to the LEP data 
is about 2 TeV and with this value, the signature of $Z'$ boson
is expected to be found via the excess in the $t \bar{t}$
pair production at the LHC or at the NLC, as is discussed
in ref. [1].

\section*{Acknowledgments}

We thank Prof. P. Ko for his valuable comments.
This work was supported in part by
the Korean Science and Engineering Foundation (KOSEF).

\end{document}